\documentclass[aps, pra, showpacs, twocolumn, %showkeys, preprintnumbers,
amssymb,superscriptaddress]{revtex4}

\usepackage{graphicx, bm, amsmath, amsfonts}

\def\ra{\rangle}
\def\la{\langle}
\def\ua{\uparrow}
\def\da{\downarrow}
\def\dag{\dagger}

\begin{document}

%\preprint{}

\title{
Entanglement Spectra of the quantum hard-square model:\\
Holographic minimal models
}
\author{Shu Tanaka}
\affiliation{
Department of Chemistry, University of Tokyo, 
7-3-1 Hongo, Bunkyo-ku, Tokyo 113-0033, Japan\\
}

\author{Ryo Tamura}
\affiliation{
International Center for Young Scientists, National Institute for Materials Science,
1-2-1, Sengen, Tsukuba-shi, Ibaraki, 305-0047, Japan
\\
}

\author{Hosho Katsura}
\affiliation{
Department of Physics, Gakushuin University, 
Mejiro, Toshima-ku, Tokyo 171-8588, Japan\\
}

\date{\today}
\begin{abstract}
We study the entanglement properties of a quantum lattice-gas model for which 
we can find the exact ground state (of the Rokhsar-Kivelson type). 
The ground state can be expressed as a superposition of states, each of 
which is characterized by a particle configuration with nearest-neighbor exclusion. 
We show that the reduced density matrix of the model on a ladder is intimately 
related to the transfer matrix of the {\it classical} hard-square model. 
The entanglement spectra of the model on square and triangular ladders 
are critical when parameters are chosen so that the corresponding classical 
hard-square models are critical. 
%correspond to the critical activities in the classical hard-square models. 
A detailed analysis reveals that the critical theories for the entanglement Hamiltonians are 
$c<1$ minimal conformal field theories. We further show that the entanglement 
Hamiltonian for the triangular ladder is {\it integrable} despite the fact that 
the original quantum lattice-gas model is non-integrable. 
\end{abstract}

% 03.67.Mn: Entanglement measures, witnesses, and other characterizations 
% 05.30.-d:	Quantum statistical mechanics
% 05.30.Rt	Quantum phase transitions (see also 64.70.Tg Quantum phase transitions 
% in specific phase transitions; and 73.43.Nq Quantum phase transitions in Quantum Hall effects)
% 32.80.Ee	Rydberg states
% 42.50.Dv Quantum state engineering and measurements in quantum optics)
% 67.85.-d	Ultracold gases, trapped gases (see also 03.75.-b Matter waves in quantum mechanics)
% 75.10.Jm: Quantized spin models, including quantum spin frustration
% 75.10.Kt:	Quantum spin liquids, valence bond phases and related phenomena

\maketitle
\section{Introduction}

Recently, there has been a growing interest in understanding the characteristics 
of quantum entanglement in many-body systems~\cite{Amico_review, Eisert_review}. 
Entanglement (von Neumann) entropy and the family of R\'enyi entropies have been 
demonstrated to be useful in quantifying bipartite entanglement in a variety of situations. 
More recently, the concept of entanglement spectrum (ES) introduced by Li and Haldane~\cite{Li_Haldane}  
has attracted considerable interest. The ES of a bipartite system consisting of 
$A$ and $B$ is obtained from the spectrum of the reduced density matrix $\rho_A$ 
for a subsystem $A$. Li and Haldane demonstrated that the entanglement spectra 
for fractional quantum Hall states reflect the gapless edge excitations at 
the {\it fictitious} boundaries created by tracing out the subsystem $B$. 
Since then, entanglement spectra have been studied in quantum Hall systems~\cite{Regnault_ES_2009, Zozulya_ES_2009, Lauchli_ES_2010, Thomale_ES_2010, Chandran_ES_2011, Qi_Katsura_Ludwig}, topological insulators~\cite{Turner_ES_2010, Fidkowski_ES, Prodan_ES_2010}, and quantum spin models 
in one~\cite{Calabrese_ES, Pollmann_ES1, Pollmann_ES2, Poilblanc_ES, Thomale_spin_chain, Laeuchli_Schliemann_ES} 
and two~\cite{Yao_Qi_ES, Cirac_et_al, Huang_ES_2011, Lou_PRB_2011} dimensions. 

In Ref.~\cite{Poilblanc_ES}, Poilblanc studied the ES of gapped two-leg spin-$1/2$ Heisenberg ladders 
and found that the ES is remarkably close to the {\it energy} spectrum of a single spin-$1/2$ Heisenberg chain, 
which is one of the subsystems of the original two-leg ladder. 
This observation was later explained analytically using first-order perturbation theory 
in the limit of strong rung coupling~\cite{Peschel_Chung_ES, Laeuchli_Schliemann_ES}. 
It is natural to expect that a similar property also holds for gapped (but non-topological) spin systems 
in two dimensions since the entanglement in those systems is short-ranged. 
From this perspective, the two-dimensional (2D) valence-bond-solid (VBS) states~\cite{Lou_PRB_2011}, 
which are exact ground states of the Affleck-Kennedy-Lieb-Tasaki (AKLT) model~\cite{AKLT, KLT}, 
and their generalizations called projected entangled pair states (PEPS)~\cite{Cirac_et_al} were examined. 
In both cases, it was found that the reduced density matrix $\rho_A$ can be interpreted as 
a thermal density matrix of a {\it holographic} (fictitious) one-dimensional (1D) system. 
In particular, for the VBS states, this holographic 1D system turns out to be a spin-$1/2$ 
Heisenberg chain.  

In this paper, we introduce and study another class of quantum many-body states, 
which falls into the category of tensor-network states~\cite{Niggemann1, Nishino-Okunishi, Hieida, Niggemann2, Martin-Delgado}. 
The quantum lattice-gas model in which these states are ground states was first constructed 
by Lesanovsky~\cite{Lesanovsky_2011}. Similar to the Rokhsar-Kivelson (RK) point in the quantum dimer model~\cite{Rokhsar-Kivelson}, the ground states are expressed as a weighted superposition of states 
labeled by classical configurations of particles. 
All the allowed configurations must satisfy the constraint that there is no more than one 
particle at any pair of neighboring lattice sites. The weight of each configuration depends on 
the parameter $z$, which corresponds to the activity (fugacity) in the classical lattice-gas model. 
The parent Hamiltonian of this state in one dimension was considered in Ref.~\cite{Lesanovsky_2011}. 
It was shown that the ground state of the 1D Rydberg lattice gas can be well approximated by this state. 
The model on the square lattice, referred to as the quantum hard-square model, was also examined 
for the validity of this ansatz~\cite{Ji_Ates_2011}. 
Interestingly, it was found that the normalization of the ground state in this case is given by 
the partition function of the {\it classical} hard-square model~\cite{Baxter_1980}. 

While the parent Hamiltonian can be easily constructed for any graph in any dimension, 
we mainly focus on the model on two-leg ladders in this paper. 
In this case, the ES can be obtained from the eigenvalues of the transfer matrix
in the classical lattice gas model in {\it two dimension}. 
From this, it immediately follows that the ES is gapless when the parameter $z$ is 
chosen so that the corresponding classical model is critical~\cite{Stephan2}.  
We numerically study the ES of our models on both square and triangular ladders at the critical point 
and find that the low-energy part of the entanglement Hamiltonian for the case of square (triangular) ladder 
is well described by the conformal field theory (CFT) with central charge $c=1/2$ ($c=4/5$). 
To our knowledge, this is the first example of quantum many-body systems whose 
entanglement Hamiltonians are precisely described by $c<1$ minimal CFTs. 
This conclusion is further supported by the nested entanglement entropy (NEE), 
the von Neumann entropy of the ground state of the entanglement Hamiltonian. 
For the model on the triangular ladder, we show that the entanglement Hamiltonian 
is {\it integrable} in the same sense as in the hard-hexagon model solved by Baxter~\cite{Baxter_1980}. 

The organization of the rest of the paper is as follows. 
In Sec. II, we define the model considered and construct the exact ground state. 
We then show that the normalization constant of this state is closely related to 
the partition function of the classical lattice-gas model in the {\it same} dimension. 
%This relation combined with Hastings' theorem allows us to infer 
%information about quantum critical points in our model.
In Sec. III, we consider our model on two-leg square and triangular ladders. 
We show how to obtain the spectrum of the reduced density matrix of the system 
from the spectrum of the two-row transfer matrix in the corresponding 2D classical model. 
Then numerical results of the entanglement entropy (EE) and correlation length are shown. 
The ES and the NEE at the critical point are studied in detail in the same section. 
%We next introduce the NEE and study its scaling properties.
%Size-dependency of the entanglement entropy at the critical point is also calculated and we confirm the area law.
We conclude with a summary in Sec. IV. 
In Appendix A, we give a detailed proof that the constructed ground state has zero-energy. 
In Appendix B, we prove the integrability of the reduced density matrix for the model 
on the triangular ladder. 

\section{The model}
\subsection{Hamiltonian}
The model we introduce is a slight generalization of the model constructed 
by Lesanovsky~\cite{Lesanovsky_2011}. It can be defined on any lattice 
in any dimension. The model describes interacting hard-core bosons on a lattice. 
The Hilbert space at each site is spanned by $|n_i\ra$, where $n_i=1$ ($0$) indicates  
that the site $i$ is occupied (empty). Note that there is at most one particle at each site. 
We further assume that the entire Hilbert space is restricted so that 
there is no more than one boson on any pair of nearest-neighboring sites. 
With the identification $|\ua\ra \leftrightarrow |1\ra$ and $|\da\ra \leftrightarrow |0\ra$, 
the operator that creates a hard-core boson at site $i$ is expressed as $\sigma^+_i {\cal P}_{\la i \ra}$, 
where $\sigma^\pm= (\sigma^x \pm i \sigma^y)/2$ with $\sigma^\alpha$ ($\alpha=x,y,z$) being  
the Pauli matrices. The projection operator $ {\cal P}_{\la i \ra}$ 
requires all sites adjacent to site $i$ to be empty. More explicitly,  
\begin{equation}
{\cal P}_{\la i \ra} = \prod_{j \in G_i} (1-n_j), 
\end{equation}
where $n_j := \sigma^+_j \sigma^-_j=(\sigma_j^z+1)/2$ and $G_i$ is a set of sites adjacent to site $i$. 
It is easy to see that $( {\cal P}_{\la i \ra})^2= {\cal P}_{\la i \ra}$. 
Note that ${\cal P}_{\la i \ra}$ commutes with any operator at site $i$. 

Let $\Lambda$ be a lattice (graph). The Hamiltonian of the model on $\Lambda$ is given by  
\begin{equation}
\label{eq:ham1}
H = \sum_{i \in \Lambda} h^\dagger_{i} (z) h_i (z),
\quad
h_i (z) = [\sigma^-_i - \sqrt{z} (1-n_i) ]  {\cal P}_{\la i \ra}.
\end{equation}
%with 
%$h_i (z) = [\sigma^-_i - \sqrt{z} (1-n_i) ]  {\cal P}_{\la i \ra}$. 
We assume that the parameter $z$ is real and nonnegative. We will see later
that $z$ exactly corresponds to the activity of the classical hard-core lattice gas model. 
To see the physical meaning of the Hamiltonian, it is illuminating to rewrite $H$ as
\begin{equation}
H= -\sqrt{z} \sum_{i \in \Lambda} (\sigma^+_i + \sigma^-_i ) {\cal P}_{\la i \ra}
+\sum_{i \in \Lambda} [(1-z) n_i+z] {\cal P}_{\la i \ra}.
\label{eq:ham2}
\end{equation}
The first term corresponds to the creation and annihilation of hard-core bosons, 
while the second term includes the chemical potential term and 
interactions among bosons that are more than two sites apart. 
For a 1D periodic chain of length $L$, the Hamiltonian reads
\begin{equation}
H = \sum^L_{i=1} {\cal P} [-\sqrt{z} \sigma^x_i + (1-3z) n_i +z n_{i-1} n_{i+1} + z] {\cal P},
\end{equation}
where ${\cal P}$ denotes the projection operator onto the physical subspace 
in which any pair of neighboring sites cannot be occupied simultaneously. 
One can interpret this Hamiltonian as the quantum Ising chain 
%where next nearest-neighbor bosons are interacted 
in a transverse and longitudinal field with nearest-neighbor exclusion and further neighbor interactions. 
In higher dimensional lattices, 
the Hamiltonian involves complicated and presumably unphysical multi-body interactions, 
e.g. $n_i n_j n_k$, and it cannot be written as the quantum Ising model 
except in special lattices such as the 3-12 (Fisher) lattice. 

\subsection{Exact ground state}
\label{sec:model}
The Hamiltonian Eq.~(\ref{eq:ham2}) in one dimension and that in 2D square lattice
were studied in Refs.~\cite{Lesanovsky_2011, Ji_Ates_2011}. It was shown that the exact ground 
state of $H$ can be obtained analytically. In addition, this exact ground state resembles 
the ground state of the Rydberg lattice gas which is experimentally realizable in cold atom systems. 
In the 1D model, explicit expressions for the energy gap and a couple of 
excited states were also obtained analytically~\cite{Lesanovsky_2012}. 

Here we consider the model on a generic lattice (graph) and find the exact ground state. 
Since each local Hamiltonian $h^\dag_i (z) h_i (z)$ is positive semi-definite, 
the Hamiltonian $H$ is also positive semi-definite and thus the energy eigenvalues 
are nonnegative. Therefore, a state with zero energy is a ground state and 
is annihilated by all $h_i (z)$. One can find that the zero-energy state takes the form
\begin{equation}
|z \ra = \frac{1}{\sqrt {\Xi (z)}} \prod_{i \in \Lambda} \exp (\sqrt{z} \sigma^+_i {\cal P}_{\la i \ra}) |\Downarrow \ra,
\label{eq:gs1}
\end{equation}
where $\Xi (z)$ is the normalization constant and $|\Downarrow\ra$ denotes the all-down state, i.e., $|\da \da ... \da \ra$, corresponding to the vacuum state. 
Note that the order of the product in Eq.~(\ref{eq:gs1}) is arbitrary 
because $[\exp(\sqrt{z} \sigma^+_i {\cal P}_{\la i \ra}), \exp(\sqrt{z} \sigma^+_j {\cal P}_{\la j \ra})]=0$ for all $i,j$. For a detailed proof of $H |z\ra =0$, see Appendix \ref{app:zero}. 
We can further show that the zero-energy state $|z\ra$ is the {\it unique} ground state of $H$. 
This can be seen by noting that all of the off-diagonal elements of $H$ are nonpositive and 
satisfy the connectivity condition, and thus the Perron-Frobenius theorem is applicable. 

Similar to the AKLT model~\cite{Arovas_Auerbach_Haldane_1988} and 
the RK model~\cite{Rokhsar-Kivelson} as well as their generalizations 
called PEPS~\cite{Verstraete_PRL_2006}, the ground state $|z \ra$ 
can be expressed as a weighted superposition of allowed states. 
To see this, it is instructive to write the unnormalized ground state
$|\Psi(z) \ra = \sqrt{\Xi (z)} |z\ra$ in terms of classical configurations: 
\begin{equation}
|\Psi(z) \ra = \sum_{\cal C \in {\cal S}} z^{n_{\cal C}/2} |{\cal C}\ra,
\label{eq:gs2}
\end{equation}
where ${\cal C}$ label classical configurations of particles on $\Lambda$ and 
${\cal S}$ is the set of configurations with nearest-neighbor exclusion. 
$|{\cal C}\ra$ correspond to the  basis states in the quantum model and 
they are orthonormal, i.e., $\la {\cal C}| {\cal C}' \ra = \delta_{{\cal C}, {\cal C}'}$. 
In the expression Eq.~(\ref{eq:gs2}), $n_{\cal C}$ counts the number of bosons in the state $|{\cal C}\ra$. 
It is now clear that the normalization constant $\Xi (z)$ is the partition function 
of the classical hard-core lattice gas model:
\begin{equation}
\Xi (z) = \la \Psi (z)| \Psi (z) \ra = \sum_{{\cal C} \in {\cal S}} z^{n_{\cal C}},
\label{eq:classical_part}
\end{equation}
and the parameter $z$ can be interpreted as an activity (fugacity) of the lattice gas. 
%One can expect that the system undergoes a phase transition from a liquid state 
%at sufficiently small $z$ to a solid state at sufficiently large $z$. 
This classical model in two dimensions has been studied extensively 
in the literature of statistical mechanics. For instance, the model on the square lattice 
is known as the hard-square model. One can easily imagine that the system undergoes 
a phase transition from a liquid state at $z \ll 1$ to a solid state at $z \gg 1$, where 
the difference of the sublattice occupations is nonzero. In fact, previous studies 
have shown that the model exhibits an order-disorder phase transition 
at $z=z_c \simeq 3.796$~\cite{Gaunt_Fisher_1965, Baxter_Enting_1980, Todo_Suzuki_1996}. 
This second-order phase transition belongs to the same universality class 
as the 2D Ising model. The model on the triangular lattice is called 
the hard-hexagon model, which is integrable and was solved by Baxter~\cite{Baxter_1980}. 
It also exhibits an order-disorder phase transition at $z=z_c=(11+5 \sqrt{5})/2=11.09017...$, 
but this transition belongs to the universality class of the three-state Potts model. 

Let us return to the original quantum model. 
A particular feature of the present construction is that correlation functions 
in the quantum ground state are the same as those of the corresponding classical model~\cite{Lesanovsky_2012}. 
This provides useful information on the ground state phase diagram of the quantum model. 
In fact, if the parameter $z$ is tuned so that the corresponding classical model is critical, then the ground state has algebraically decaying correlation functions, which suggests that the quantum model is critical as well.
This type of quantum critical point is called {\it conformal quantum critical point} 
because the ground state wave functional is conformally invariant in the scaling limit~\cite{Ardonne_Fendley_2004}. 

\section{Entanglement properties}
\begin{figure}
%\centerline{\includegraphics[width=.9\columnwidth, clip]
%7cm,clip]
%{figures/sblock_spec_combine_2.eps}}
\begin{center}
\includegraphics{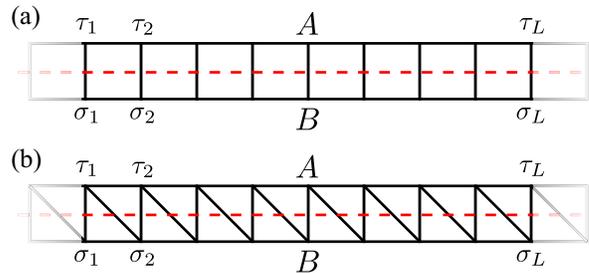}
\end{center}
\caption{
(color online)
(a) Two-leg square ladder and (b) two-leg triangular (zigzag) ladder. 
The systems are divided into two parts, $A$ and $B$, as indicated by the dashed lines.
We impose periodic boundary conditions in the leg direction: $\tau_{L+1}=\tau_1$ and $\sigma_{L+1}=\sigma_1$. 
}
\label{fig:ladder}
\end{figure}

In this section, we consider the exact ground state Eq.~(\ref{eq:gs1}) on a two-leg ladder 
as shown in Fig. \ref{fig:ladder}. The classical partition function Eq.~(\ref{eq:classical_part}) 
is defined on the same ladder, which is still 1D. 
This implies that there is no critical point at finite $z$ 
and the correlation functions of local operators in the ground state are exponentially decaying. 
Nevertheless, we find an intimate relation to CFTs, which
will be revealed by the analysis of the entanglement properties. 
We show that the ES of this system can be inferred from the spectrum 
of the transfer matrix in the corresponding 2D classical system. 
We also study the von Neumann entropy associated with 
the ground state of the entanglement Hamiltonian, and show that the underlying CFT is 
a unitary minimal model with central charge $c<1$. 

%--------------------------------------------------------------------------
%--------------------------------------------------------------------------

%--------------------------------------------------------------------------
\begin{figure}
%\centerline{\includegraphics[width=.9\columnwidth, clip]
%7cm,clip]
%{figures/sblock_spec_combine_2.eps}}
\begin{center}
\includegraphics{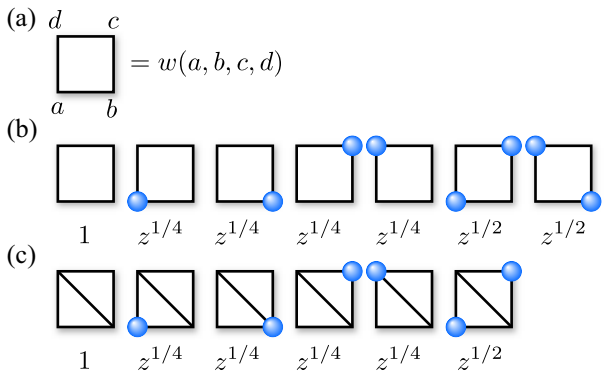}
\end{center}
\caption{
(color online)
(a) Graphical representation of the local Boltzmann weights. 
 Allowed configurations and their Boltzmann weights for the square ladder (b) and the triangular ladder (c). 
}
\label{fig:weights}
\end{figure}
%--------------------------------------------------------------------------
 
\subsection{Reduced density matrix}
\label{sec:RDM}
Consider the two-leg ladder of length $L$ with periodic boundary conditions in the leg direction. 
We divide the system into two subsystems, $A$ and $B$ as shown in Fig. \ref{fig:ladder}. 
Let $\tau=\{ \tau_1, \tau_2, ..., \tau_L \}$ be a particle configuration on the chain in $A$, 
and  $\sigma=\{ \sigma_1, \sigma_2, ..., \sigma_L \}$ be that on the chain in $B$. 
Here $\tau_i=1$ ($0$) if the site $i$ on $A$ is occupied (empty). The same applies to $\sigma_i$. 
Let $|\tau \ra$ and $|\sigma \ra$ denote the corresponding basis states in the quantum model. 
Then the (unnormalized) ground state $|\Psi (z) \ra$ on the ladder is written as 
\begin{equation}
|\Psi (z) \ra = \sum_{\tau} \sum_{\sigma} [T (z)]_{\tau, \sigma} |\tau \ra \otimes |\sigma \ra, 
\label{eq:gs3}
\end{equation}
where 
\begin{equation}
[T(z)]_{\tau, \sigma} := \prod^L_{i=1} w(\sigma_i, \sigma_{i+1}, \tau_{i+1}, \tau_{i}), 
\label{eq:tmat1}
\end{equation}
with $w (a,b,c,d)$ being the Boltzmann weights for each face (see Fig. \ref{fig:weights}). 
More explicitly, for the square ladder, we have 
\begin{equation}
[T (z) ]_{\tau, \sigma} = \prod^L_{i=1} z^{(\sigma_i + \tau_i)/2} (1-\sigma_i \tau_i) 
(1-\sigma_i \sigma_{i+1}) (1-\tau_i \tau_{i+1}), 
\end{equation}
and for the triangular ladder, we have
\begin{eqnarray}
 \nonumber
 [T (z) ]_{\tau, \sigma} &=&\prod^L_{i=1} z^{(\sigma_i + \tau_i)/2}
 (1-\sigma_i \tau_i) (1-\sigma_i \sigma_{i+1})\\
 &\times &(1-\tau_i \tau_{i+1})(1-\tau_i \sigma_{i+1}). 
\end{eqnarray}
One can think of $T(z)$ in Eq.~(\ref{eq:gs3}) as an $N_L$-dimensional matrix,  
where $N_L$ is the number of allowed configurations in each chain~\cite{Fibonacci_num}. 
In fact, we can identify $T(z)$ as the transfer matrix of the 2D classical lattice gas model with hard-core exclusion~\cite{Baxter_1980}. 
%As we will see in the following, this correspondence plays a crucial role in studying 
%the entanglement spectra of our model. 
Note that a similar transfer matrix formalism was also applied 
to entanglement entropies of the 2D RK wavefunction~\cite{Stephan1}.

The reduced density matrix for $A$ describing the entanglement between the two subsystems 
is defined by $\rho_{A} := {\rm Tr}_{B} [| z \ra \la z|]$. 
%\textbf{[H.K.: We should decide later whether $|\Psi(z)\ra$ or $|z\ra$ is suitable here.]} 
To obtain the spectrum of $\rho_{A}$, we follow the approach used in Ref.~\cite{Jozsa_PRA_2000}. 
A similar approach has been applied to 
%the reduced density matrix of 
the 2D VBS states~\cite{KKKKT, Lou_PRB_2011}. 
We first write the state $|z \ra$ as
\begin{equation}
|z \ra = \frac{1}{\sqrt{\Xi (z)}} \sum_{\sigma} |\varphi_\sigma \ra \otimes | \sigma \ra,
\end{equation}
where $|\varphi_\sigma \ra := \sum_\tau [T(z)]_{\tau, \sigma} |\tau \ra$. 
It should be noted that the basis states $|\varphi_\sigma \ra$ are not orthonormal. 
Since the basis states in $B$, i.e., $|\sigma \ra$, are orthonormal, one can easily 
trace out the degrees of freedom in $B$ and obtain
\begin{equation}
\rho_A = \frac{1}{\Xi (z)} \sum_\sigma |\varphi_\sigma \ra \la \varphi_\sigma |.
\end{equation}
We next introduce the Gram matrix $M$ (overlap matrix) whose matrix elements are given by
\begin{equation}
M_{\sigma', \sigma} = \frac{1}{\Xi (z)} \la \varphi_{\sigma'}| \varphi_{\sigma} \ra. 
\end{equation}
Since the states $|\tau \ra$ (in $A$) are also orthonormal, one finds
\begin{equation}
M = \frac{1}{\Xi (z)} [T(z)]^{\rm T} T(z),
\label{eq:Gram1}
\end{equation}
where the superscript ${\rm T}$ denotes matrix transpose. 
From the argument in~\cite{Jozsa_PRA_2000}, we can show the following 
properties: (i) ${\rm Tr}[M]=1$, (ii) all the eigenvalues of $M$ are nonnegative, 
and (iii) nonzero eigenvalues of $M$ and $\rho_A$ are identical. 
It follows from (iii) that the ES associated with $\rho_A$ is 
identical to the eigenvalue spectrum of the matrix $M$. 
%\textbf{[H.K.: Should we write proofs of these properties? Maybe we can write a general results in Appendix B.]}
The entanglement Hamiltonian of our model is then defined via $M := \exp (-H_{\rm E})$. 
One can regard $M$ as a thermal density matrix of an auxiliary 1D model, which we call the {\it holographic model}. 

\begin{figure}
%\centerline{\includegraphics[width=.9\columnwidth, clip]
%7cm,clip]
%{figures/sblock_spec_combine_2.eps}}
\includegraphics{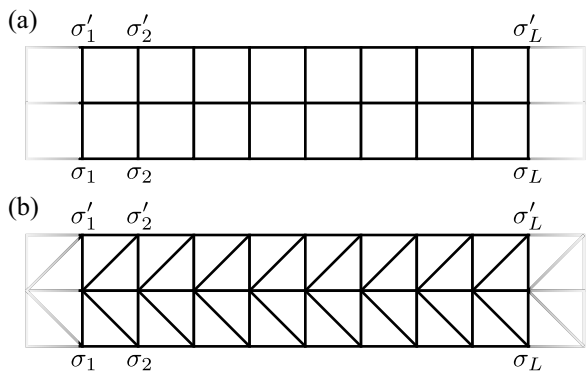}
\caption{
Two-row transfer matrix for the square ladder (a), and that for the triangular ladder (b).
}
\label{fig:two-row}
\end{figure}

Let us now discuss the relation between the spectrum of $M$ and that of the transfer matrix $T(z)$. 
For the square ladder, we have $w (\sigma_i, \sigma_{i+1}, \tau_{i+1}, \tau_i) = w (\tau_i, \tau_{i+1}, \sigma_{i+1}, \sigma_i)$ 
which yields $[T (z)]^{\rm T} = T (z) $. We thus obtain $M = [T(z)]^2/ \Xi (z)$. 
The matrix $[T(z)]^2$ can be interpreted as a two-row transfer matrix 
by which the classical hard-square model is described (see Fig. \ref{fig:two-row} (a)). 
For the triangular ladder, the local Boltzmann weights do not have the above symmetry. 
Instead, we can interpret the matrix $[T(z)]^{\rm T} T(z)$ as a product of transfer matrices 
defined on the three-leg ladder shown in Fig. \ref{fig:two-row} (b). One can view the lattice as 
a triangular lattice deformed into a topologically equivalent square lattice with diagonal bonds. 
Therefore, the two-dimensinal classical model corresponding to our quantum model 
on the triangular ladder is the hard-hexagon model. 
For both square and triangular ladders, the entanglement spectra can be read off 
from the spectra of transfer matrices~\cite{GGE}. 
From this correspondence, it is clear that the ES is critical if the activity $z$ 
is chosen so that the corresponding classical model is critical. 
In Sec. \ref{sec:ES}, we study numerically the spectrum of $M$ and confirm that this is indeed the case. 

%--------------------------------------------------------------------------
%--------------------------------------------------------------------------

\subsection{Entanglement entropy (EE)}

\begin{figure}
%\centerline{\includegraphics[width=.9\columnwidth, clip]
%7cm,clip]
%{figures/sblock_spec_combine_2.eps}}
\includegraphics[width=8.5cm]{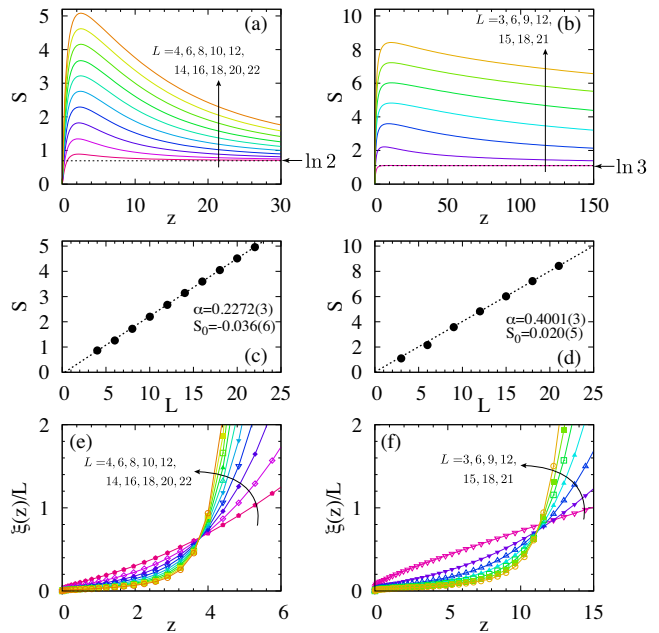}
\caption{
(color online)
(a) EE ($S$) of the state Eq.~(\ref{eq:gs3}) on the square ladder 
as a function of activity $z$ and (b) the same for the triangular ladder. 
As indicated by the arrows, $S$ increases linearly with increasing system size $L$ in both cases. 
%EE of our model on square ladder (a) and triangular ladder (b) as functions of $z$. 
The dotted horizontal lines indicate, respectively, $\ln 2$ and $\ln 3$ which are the EEs at $z=\infty$. 
(c) Size-dependence of $S$ at the critical activity $z=z_c$ for the square ladder with different sizes $L=4-22$ 
and (d) the same for the triangular ladder with $L=3-21$. 
The dotted lines are least squares fits to the last four data points using Eq.~(\ref{eq:area_law}). 
The fitting parameters are shown in the figures. 
(e) Correlation length divided by system size $\xi(z)/L$ versus $z$ for the square ladder and (f) that for the triangular ladder. 
}
\label{graph:EEvsZ-ladder}
\end{figure}

We now study the EE which can be obtained from 
\begin{eqnarray}
S=-{\rm Tr}[ M \ln M] =-\sum_{\alpha}p_{\alpha}\ln p_{\alpha}, 
\label{entropy}
\end{eqnarray}
where $p_{\alpha}$ ($\alpha=1,2,\cdots,N_L$) are the eigenvalues of $M$.  
We calculate EE using exact diagonalization for the square and triangular ladders with periodic boundary conditions (see Fig. \ref{fig:ladder}). 
In order to avoid boundary conditions incompatible with the modulation of the densest packings, 
%eliminate boundary effects, 
we assume that $L=2m$ and $L=3m$ 
($m \in \mathbb{N}$) for the square and triangular ladders, respectively.  
The dependence of EE on the activity $z$ for both models is 
%the square ladder and that for the triangular ladder are 
shown in Figs. \ref{graph:EEvsZ-ladder} (a) and \ref{graph:EEvsZ-ladder} (b). 
In the limit $z \to \infty$, the EEs for square and triangular ladders become $\ln 2$ and $\ln 3$, respectively, 
irrespective of the system size $L$. 
%, corresponding to the degree of freedom in $A$.
This can be understood as follows: for large $z$, the ground state is approximately given by a superposition of the ordered states with the maximum density of particles. 
These ordered states are related to each other by translations. 
Therefore, we have $\rho_A \sim \frac{1}{\sqrt 2}(|0101...\ra \la 0101 ...|+|1010...\ra \la 1010...|)$ for the square ladder and a similar one with period 3 for the triangular ladder. 
As a result, we obtain the observed saturation values of EE. 
%In this limit, the ground state is fully occupied state so that adjacent sites of occupied sites are empty.
In the opposite limit $z \to 0$, the EEs become zero since the ground state is the vacuum state (see Eq.~(\ref{eq:gs2})). 
In the intermediate region between these two limits, the EE shows a non-monotonic dependence on $z$ 
and has a peak in both square and triangular ladders as shown in Figs.~\ref{graph:EEvsZ-ladder} (a) and \ref{graph:EEvsZ-ladder} (b). In both cases, the peak position is at about $z=z_c$, i.e., the critical activity in the corresponding  classical model, and remains almost unchanged with increasing $L$. 

Let us focus on the entanglement properties of the model at $z=z_c$. 
Figures \ref{graph:EEvsZ-ladder} (c) and \ref{graph:EEvsZ-ladder} (d) show the scaling of 
the EE $S(L)$ for both the square and triangular ladders. 
From these plots, it is clear that the EEs at $z=z_c$ scale linearly with the system size $L$ 
(corresponding to the length of the boundary between $A$ and $B$) and thus obey the area law:
\begin{eqnarray}
 \label{eq:area_law}
 S(L) = \alpha L + S_0,
\end{eqnarray}
where $\alpha$ and $S_0$ are the fitting parameters independent of $L$. 
For square and triangular ladders, $(\alpha,S_0)=(0.2272(3),-0.036(6))$ and $(\alpha,S_0)=(0.4001(3),0.020(5))$, respectively.
In both cases, $S_0$ is nearly zero, suggesting that the topological EE 
introduced in Refs.~\cite{Kitaev-Preskill, Levin-Wen} is zero in our system. 
This is consistent with the fact that CFTs describing 
the entanglement Hamiltonians of our system are {\it non-chiral} as we will see later.  

%--------------------------------------------------------------------------
%--------------------------------------------------------------------------

\begin{figure}
%\centerline{\includegraphics[width=.9\columnwidth, clip]
%7cm,clip]
%{figures/sblock_spec_combine_2.eps}}
\includegraphics[width=8.5cm]{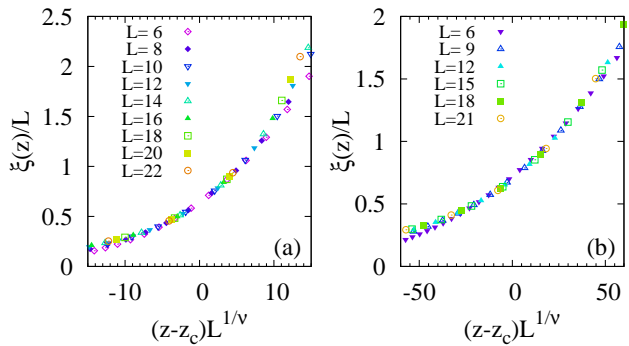}
\caption{
(color online)
Finite-size scaling plot of $\xi(z)/L$ for the square ladder (a) 
and that for the triangular ladder (b). The correlation length exponents 
of the square and triangular ladders are $\nu=1$ and $\nu=5/6$, respectively.
}
\label{graph:FSS-ladder}
\end{figure}

%the second-order phase transition should occur at $z=z_c$ and the EEs should diverge at $z=z_c$ in the thermodynamic limit. 
As discussed in the previous subsection, the Gram matrix $M$ can be interpreted as a transfer matrix
in the corresponding 2D classical model. Therefore, we expect that the anomalous behavior of EE at $z=z_c$ is attributed to the phase transition in the classical model. 
To study the nature of the phase transitions in both square and triangular cases, 
%Next, in order to confirm the universality class of the phase transitions at $z=z_c$, 
we perform a finite-size scaling analysis of the correlation length. 
The correlation length is defined in terms of the entanglement gap as 
\begin{eqnarray}
 \xi(z) := \frac{1}{\ln [p^{(1)}(z)/p^{(2)}(z)]},
\end{eqnarray}
where $p^{(1)}(z)$ and $p^{(2)}(z)$ are the largest and the second-largest eigenvalues of $M$ at $z$, respectively. 
Figures \ref{graph:EEvsZ-ladder} (e) and \ref{graph:EEvsZ-ladder} (f) show the correlation length 
divided by the system size $L$ as a function of $z$ for both square and triangular cases.  
Clearly, the curves for different system sizes cross at the same point $z=z_c$, 
which implies that the dynamical critical exponent is given by $1$.  
Near the critical point, we expect that $\xi(z)/L$ obeys the scaling relation 
%Here we perform the finite-size scaling plots of the correlation length using the following equation:
%
\begin{eqnarray}
 \xi(z)/L = f((z-z_c)L^{1/\nu}),
\end{eqnarray}
where $\nu$ is the correlation length exponent and $f(\cdot)$ is a scaling function. 
Figures \ref{graph:FSS-ladder} (a) and \ref{graph:FSS-ladder} (b) show plots of $\xi(z)/L$ versus the scaling variable $(z-z_c)L^{1/\nu}$. 
For the square ladder, we find a good data collapse with $\nu=1$, which agrees with the correlation length exponent of the 2D Ising model  (see Fig. \ref{graph:FSS-ladder} (a)). 
On the other hand, for the triangular ladder, the exact value $\nu=5/6$ can be obtained by noting 
that the two-row transfer matrix shown in Fig. \ref{fig:two-row} (b) is exactly equivalent to that of the hard-hexagon model~\cite{Baxter_Pearce_1982}. 
We obtain an excellent data collapse as shown in Fig. \ref{graph:FSS-ladder} (b). 
Note that the exponent $\nu=5/6$ coincides with that of the three-state Potts model~\cite{Wu_RMP_1982}. 
%using $\nu=6/5$, which is inferred from the known exact value of the three-state Potts model (Wu's paper (p. 261))

%--------------------------------------------------------------------------
%--------------------------------------------------------------------------

\subsection{Entanglement spectrum (ES)}
\label{sec:ES}
\begin{figure}
%\centerline{\includegraphics[width=.9\columnwidth, clip]
%7cm,clip]
%{figures/sblock_spec_combine_2.eps}}
\includegraphics[width=8.5cm]{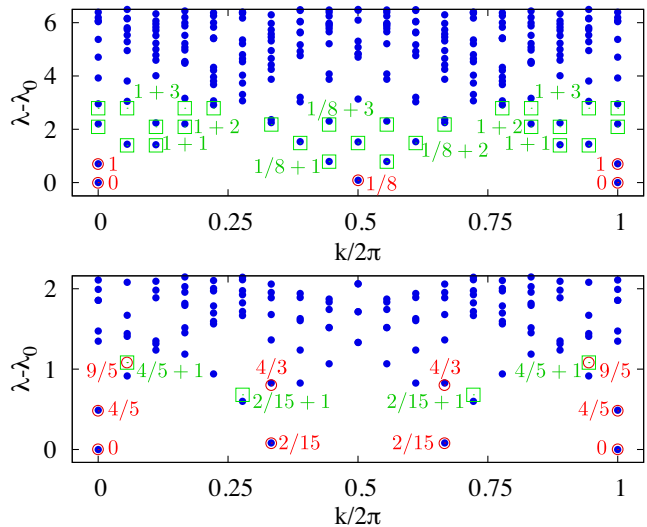}
\caption{
(color online)
Low-energy spectra of the entanglement Hamiltonian $H_{\rm E}$ at $z=z_c$
for the square ladder (top panel) and for the triangular ladder (bottom panel). 
In both cases, the system size is $L=18$. 
The ground state energy of $H_{\rm E}$ is denoted by $\lambda_0$. 
The red open circles indicate positions of the primary fields of the corresponding CFTs, 
while the green open squares imply the positions of the descendant fields. 
Discrepancies between the numerical results and the CFT predictions are due to finite-size effects.
}
\label{graph:ES-ladder}
\end{figure}

The anomalous behavior of the correlation length at $z=z_c$ suggests that the entanglement gap 
at this point vanishes linearly with $1/L$. To further elucidate the gapless nature of the entanglement 
Hamiltonian $H_{\rm E}:=-\ln M$, 
%at the critical point, 
we calculate the excitation spectrum of $H_{\rm E}$ at the critical point. 
The ES $\{ \lambda_\alpha \}_{\alpha=1,...,N_L}$ can be obtained from the relation $\lambda_\alpha = -\ln p_\alpha$. Each eigenstate is labeled by the total momentum $k$ due to the translational symmetry in the leg direction. 
Fig. \ref{graph:ES-ladder} shows the ES of both square and triangular ladders at $z=z_c$. 
In both cases, there is the minimum eigenvalue $\lambda_0(:=\min_\alpha \lambda_\alpha)$ at $k=0$, and the ES is symmetric about $k=0$ and $k=\pi$ (mod $2\pi$). 
%square and triangular ladders for $L=18$ at $z=z_c$. 
For the square ladder, the gapless modes at momenta $k=0$ and $k=\pi$ are clearly visible. 
On the other hand, for the triangular ladder, they are at $k=0$, $2\pi/3$, and $4\pi/3$. 

These towers of energy levels can be identified as those of CFTs describing the low-energy spectra of $H_{\rm E}$. As opposed to topologically ordered systems such as fractional quantum Hall states, 
the underlying CFTs are non-chiral in our case, i.e., there are both left- and right-moving modes. 
In general, the excitation energies from which towers are generated have the form
\begin{equation}
\lambda_\alpha - \lambda_0 = \frac{2\pi v}{L} (h_{L,\alpha}+h_{R,\alpha}),
\label{eq:CFT_spec}
\end{equation} 
where $v$ is the velocity, and $h_{L, \alpha}$ and $h_{R,\alpha}$ are (holomorphic and anti-holomorphic) 
conformal weights. The sum $h_{L, \alpha}+h_{R,\alpha}$ is called scaling dimension. 
$h_{L,\alpha}-h_{R,\alpha}$ is related to the total momentum $k$. 
Comparing the energy spectrum obtained numerically with the above relation, 
we can identify conformal weights of low-lying states. 
Scaling dimensions of several low-lying states are indicated in Fig. \ref{graph:ES-ladder}. 
From the obtained scaling dimensions, we conclude that low-energy spectrum of $H_{\rm E}$ 
for the square ladder model at $z=z_c$ is described by the CFT with central charge $c=1/2$~\cite{Henkel_text}, 
whereas that for the triangular ladder model is described by the $c=4/5$ CFT. 
The former CFT is identical to that of the 2D critical Ising model, 
while the latter describes the universality class of the critical three-state Potts model. 
These are all consistent with the analysis of correlation length exponent in the previous subsection. 

We note that the exact scaling dimensions for the triangular ladder model at the critical point 
can be obtained analytically by exploiting the integrability of the hard-hexagon model~\cite{Klumper-Pearce}. 
Here integrability means the existence of a family of commuting transfer matrices. 
In fact, for the model on the triangular ladder, we can show that there is a one-parameter family of 
commuting matrices including the Gram matrix $M$ in Eq.~(\ref{eq:Gram1}) as a special case 
(see Appendix \ref{app:commuting} for the proof). This property holds even away from the critical point. 
Therefore, the entanglement Hamiltonian of this system at arbitrary $z$ is integrable 
in the same sense as in the original hard-hexagon model. 

%--------------------------------------------------------------------------
%--------------------------------------------------------------------------

\subsection{Nested entanglement entropy (NEE)}

To provide further evidence that the entanglement Hamiltonians at $z=z_c$ are described by minimal CFTs with central charge $c<1$, we next consider NEE which was first introduced in Ref.~\cite{Lou_PRB_2011}. 
The NEE is the entanglement (von Neumann) entropy associated with the ground state of $H_{\rm E}$. 
One might think that the scaling analysis presented in the previous subsection suffices to find the central charges. 
However, the NEE provides a simpler and more clear-cut approach. 
As discussed in Sec. \ref{sec:RDM}, the Gram matrix $M=\exp (-H_{\rm E})$ can be interpreted as a thermal density matrix of the holographic model in one dimension. 
This implies that $H_{\rm E}$ can be written as $H_{\rm E} = \beta_{\rm eff} H_{\rm hol}$ with $H_{\rm hol}$ being the Hamiltonian for the holographic model. Here $\beta_{\rm eff}$ is the fictitious inverse temperature and is not universal. 
We are interested in $H_{\rm hol}$ rather than $H_{\rm E}$. However, there is no unique way to disentangle $\beta_{\rm eff}$ from $H_{\rm hol}$. To overcome this, we first note that the ground state of $H_{\rm E}$ is the same as that of $H_{\rm hol}$, corresponding to the eigenvector of $M$ associated with the largest eigenvalue. 
We then recall that for 1D critical systems, one can read off the underlying CFT solely from the entanglement entropy obtained from the ground state of $H_{\rm hol}$ \cite{Calabrese_Cardy}. 
Therefore, one can obtain the central charge directly from the scaling properties of the NEE without evaluating
$\beta_{\rm eff}$ or the velocity $v$ in Eq.~(\ref{eq:CFT_spec}), as demonstrated in the previous work \cite{Lou_PRB_2011}. 

Let us now give the definition of the NEE. We divide the system on which the holographic model lives into two subsystems: 
a block of $\ell$ consecutive sites and the rest of the chain. 
The nested reduced density matrix is then defined as
\begin{eqnarray}
 \rho(\ell) := {\rm Tr}_{\ell+1,\cdots,L} [ | \psi_0 \ra \la \psi_0 | ],
\end{eqnarray}
where $|\psi_0\ra$ denotes the normalized ground state of $H_{\rm E}$ and 
the trace is taken over the degrees of freedom outside the block. 
Using $\rho (\ell)$, the NEE is expressed as
\begin{eqnarray}
 s(\ell,L) := -{\rm Tr}_{1,\cdots,\ell} [ \rho(\ell) \ln \rho(\ell)],
\end{eqnarray}
where the trace is taken over the states in the block. 

\begin{figure}
%\centerline{\includegraphics[width=.9\columnwidth, clip]
%7cm,clip]
%{figures/sblock_spec_combine_2.eps}}
\includegraphics[width=8.5cm]{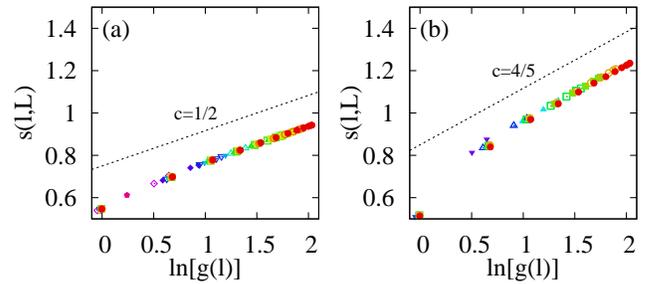}
\caption{
(color online)
NEE for the square ladder (a) and that for the triangular ladder (b). 
%as a functions of $\ln[g(\ell)]$. 
Dotted lines are a guide to the eye and indicate the slope of $c/3$, where $c=1/2$ and $c=4/5$ are used for the square and triangular ladders, respectively. 
The solid circles indicate NEE for $L=24$. 
The other symbols are the same as in Fig. ~\ref{graph:FSS-ladder}. 
}
\label{graph:nee-ladder}
\end{figure}

Let us now analyze the numerically obtained NEE in detail. 
Since the low-energy spectra of $H_{\rm E}$ at $z=z_c$ show good agreement 
with those of CFT predictions, the NEE is expected to behave as
\begin{eqnarray}
\label{eq:CC_formula}
 &&s(\ell,L) = \frac{c}{3}\ln[g(\ell)] + s_1,\\
 &&g(\ell) = \frac{L}{\pi} \sin \left( \frac{\pi \ell}{L} \right),
\end{eqnarray}
where $s_1$ is a non-universal constant~\cite{Calabrese_Cardy}. 
Figure \ref{graph:nee-ladder} shows the NEE for both square and triangular ladders as a function of $\ln[g(\ell)]$. The data for $L=24$ are obtained by the power method. 
The slopes of the dotted lines in Fig.~\ref{graph:nee-ladder} are $c/3$, where $c=1/2$ for the square ladder and $c=4/5$ for the triangular ladder. 
Clearly, the results are in excellent agreement with the formula Eq.~(\ref{eq:CC_formula}) 
and provide further evidence for the holographic minimal models, 
i.e., the entanglement Hamiltonians associated with the square and triangular ladders at $z=z_c$ 
are described by unitary minimal CFTs with $c<1$. 

%--------------------------------------------------------------------------
%--------------------------------------------------------------------------

\section{Conclusion}
We have presented a detailed analysis of the entanglement entropy and entanglement spectrum (ES) 
in the ground state of the quantum lattice-gas model on two-leg ladders. 
The exact ground state of the model can be obtained as a superposition of states, 
each of which is labeled by a classical configuration with nearest neighbor exclusion. 
We have shown that the reduced density matrix of one of the legs can be written 
in terms of the transfer matrix in the classical lattice-gas model in two dimensions. 
We then numerically studied the entanglement properties of the models on both square and triangular ladders. 
Both models exhibit critical ES when the parameter (activity) 
$z$ is chosen so that the corresponding classical model is critical. 
From the finite-size scaling analysis, we found that the critical theory describing 
the gapless ES of the square ladder is the CFT with $c=1/2$, while that of the triangular ladder is the CFT with $c=4/5$. 
This was further confirmed by the analysis of the nested entanglement entropy. 
We also showed that the model on the triangular ladder is integrable at arbitrary $z$ 
in the sense that there is a one-parameter family of matrices commuting with the reduced density matrix. 

It would be interesting to calculate other quantities by exploiting the close connection 
between the entanglement Hamiltonian and the two-dimensional classical model. 
It is likely that the virtual-space transfer matrix method~\cite{Suzuki_1985} can be applied 
to the calculation of R\'enyi entropies. 
%In particular, the single-copy entanglement \cite{Eisert_Cramer_2005}, a special case of the R\'enyi entropy, 
%can be obtained from the largest eigenvalue of the transfer matrix and the partition function. 
Finally, it would also be interesting to extend our analysis to the case of truly two-dimensional lattices. 
We note, however, that the model on $N$-leg ladders ($N>2$) cannot be treated 
on the same footing since the Gram matrix cannot be written by the transfer matrix alone.  
Thus new methods need to be developed to analyze two-dimensional systems. 

%%%%%%%%%%%%%%%%%%%%%%%%%%%%%%%%%%%%%%%
%%%%%%       ACKNOWLEDGMENT       %%%%%%
%%%%%%%%%%%%%%%%%%%%%%%%%%%%%%%%%%%%%%%
\section*{Acknowledgment}
The authors thank Shunsuke Furukawa, Tohru Koma, and Tsubasa Ichikawa for valuable discussions. 
This work was supported by Grant-in-Aid for JSPS Fellows (23-7601) and for Young Scientists (B) (23740298). 
R.T. is partly supported financially by National Institute for Materials Science (NIMS).
Numerical calculations were performed on supercomputers at the Institute for Solid State Physics, University of Tokyo.

\appendix
\section{Zero-energy ground state}
\label{app:zero}
Here we prove that the state $|z\ra$ defined in Eq.~(\ref{eq:gs1}) is the zero-energy 
ground state of the Hamiltonian. Since the Hamiltonian takes the form 
$H=\sum_{i \in \Lambda} h^\dagger_i (z) h_i (z)$, it suffices to see that 
$|z \ra$ is annihilated by all $h_i (z)$. For notational convenience, we introduce the operator 
$g^\dagger_i (z) := \exp (\sqrt{z} \sigma^+_i {\cal P}_{\la i \ra}) = 1 + \sqrt{z} \sigma^+_i {\cal P}_{\la i \ra}$. 
Recalling that $[g^\dagger_i (z), g^\dagger_j (z') ] =0$ $\forall i,j$ and $(1-n_i) \sigma^+_i =0$, we have
\begin{eqnarray}
h_i (z) |\Psi (z) \ra &=& h_i (z) g^\dagger_i (z) \prod_{j  \in \Lambda \backslash \{ i \}} g^\dagger_j (z) |\Downarrow \ra 
\nonumber \\
&=& {\cal P}_{\la i \ra} \sigma^-_i \prod_{j  \in \Lambda \backslash \{ i \}} g^\dagger_j (z) |\Downarrow \ra. 
\end{eqnarray}
We can further simplify the above expression by noting that $\sigma^-_i$ commutes with $g^\dagger_j (z)$ 
when $i$ and $j$ are not adjacent, and $\sigma^-_i g^\dagger_j (z) = \sigma^-_i$ when $i$ and $j$ are adjacent. 
It then follows from $\sigma^-_i |\Downarrow \ra =0$ that $h_i (z) |\Psi (z) \ra =0$ for any $i \in \Lambda$, which proves that $|z \ra$ is a zero-energy 
ground state of $H$. For the uniqueness of the ground state, see the discussion in Sec. \ref{sec:model}. 

\section{Hard-square model with diagonal interactions}
\label{app:commuting}
To find a one-parameter family of commuting matrices for the model on the triangular ladder, 
we consider a larger class of models in which the hard-square and hard-hexagon models are 
included as limiting cases. This generalized model was first introduced by Baxter to solve 
the hard-hexagon model~\cite{Baxter_1980}. 

\begin{figure}[t]
%\centerline{\includegraphics[width=.9\columnwidth, clip]
%7cm,clip]
%{figures/sblock_spec_combine_2.eps}}
%\includegraphics[width=8.5cm]{}
\includegraphics{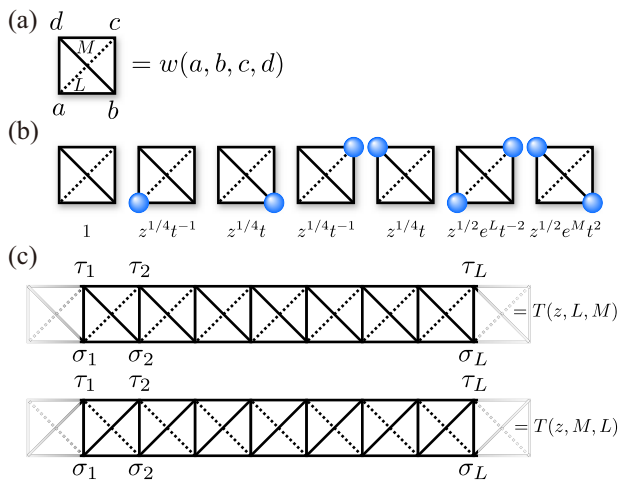}
\caption{
(a) Graphical representation of the local Boltzmann weights. 
(b) Allowed configurations and their Boltzmann weights. 
(c) Row-to-row transfer matrix $T(z,L,M)$ and its transpose $[T(z,L,M)]^{\rm T} =T(z,M,L).$
}
\label{fig:diagonal}
\end{figure}

Besides the activity $z$, there are two interaction parameters $L$ and $M$ (not to be confused with the system size $L$ and the Gram matrix $M$ as used in the main text) in the generalized model. 
The transfer matrix of the system corresponding to Eq.~(\ref{eq:tmat1}) is defined as
\begin{equation}
[T(z,L,M)]_{\tau, \sigma} := \prod_{i} w (\sigma_i, \sigma_{i+1}, \tau_{i+1}, \tau_i)
\end{equation}
with 
\begin{eqnarray}
w (a,b,c,d) &=& z^{(a+b+c+d)/4} e^{L ac+M bd} t^{-a+b-c+d} \nonumber \\
&\times & (1-ab) (1-bc) (1-cd) (1-da),
\label{eq:BW_gen}
\end{eqnarray}
where $a,b,c,d$ take the value $0$ or $1$. 
Here we again consider the case with periodic boundary conditions in the horizontal direction. 
The parameter $t$ cancels out in $T(z,L,M)$ and so is arbitrary. 
$L$ and $M$ represent the diagonal interactions as shown in Fig. \ref{fig:diagonal} (a). 
The Boltzmann weights for the allowed configurations of each face are shown in Fig. \ref{fig:diagonal} (b). 
The model on the square ladder is obtained by setting $L=M=0$, 
while the model on the triangular ladder is obtained by taking the limit $L=0$ and $M \to -\infty$. 

Baxter showed that the statistical mechanics model with Boltzmann weights Eq.~(\ref{eq:BW_gen}) is exactly solvable if $z, L, M$ satisfy 
\begin{equation}
z = \frac{(1-e^{-L})(1-e^{-M})}{e^{L+M}-e^L-e^M}. 
\label{eq:manifold}
\end{equation}
He also showed that two transfer matrices $T(z,L,M)$ and $T(z', L', M')$ commute if 
$z,L,M$ and $z',L',M'$ satisfy the above condition and 
\begin{equation}
z^{-1/2} (1-z e^{L+M}) = (z')^{-1/2} (1-z' e^{L'+M'}). 
\label{eq:delta}
\end{equation}
On the integrable manifold represented by Eq.~(\ref{eq:manifold}), the activity $z$ is determined 
as a function of $L$ and $M$. However, it is unconstrained in the special limit $L=0$ and $M \to -\infty$, 
corresponding to the hard-hexagon model. Note that the pure hard-square model is not 
solvable since Eq.~(\ref{eq:manifold}) has only the trivial solution $z=0$ if $L=M=0$. 

%--------------------------------------------------------------------------
%--------------------------------------------------------------------------

Let us return to our model on the two-leg ladder. The Gram matrix in Eq.~(\ref{eq:Gram1}) can be embedded 
in the following larger class of matrices:
\begin{equation}
M(z,L,M) := \frac{1}{\Xi (z,L,M)} [T(z,L,M)]^{\rm T} T(z,L,M),
\end{equation}
where $\Xi (z,L,M)={\rm Tr}[T(z,L,M)]^{\rm T} T(z,L,M)$. 
Since the relations (\ref{eq:manifold}) and (\ref{eq:delta}) are symmetric with respect to 
the interchange of $L$ and $M$, we have $[ T(z,L,M), T(z,M,L) ]=0$. 
By the same reasoning, $[T(z, L, M), T(z', M', L')]=0$ if $z', L', M'$ satisfy both Eqs. (\ref{eq:manifold}) and (\ref{eq:delta}). 
On the other hand, from Eq.~(\ref{eq:BW_gen}), it is easy to see that $[T(z,L,M)]^{\rm T} = T(z,M,L)$ 
(see Fig. \ref{fig:diagonal} (c)). It immediately follows that the matrix $T(z, L, M)$ commutes with its transpose. 
Therefore, four matrices $T(z, L, M)$, $[T(z, L, M)]^{\rm T}$, $T(z', L', M')$, and $[T(z', L', M')]^{\rm T}$ 
are mutually commuting if $z, L, M$ and $z', L', M'$ satisfy both the condition (\ref{eq:manifold}) 
and the relation (\ref{eq:delta}). This ensures the existence of 
the family of commuting matrices $M(z,L,M)$. The matrix specializes to the Gram matrix for 
the triangular ladder model when $L=0$ and $M \to -\infty$.


\begin{thebibliography}{99}
\bibitem{Fibonacci_num}
For the periodic ladder, $N_L = F_{L-1}+F_{L+1}$, where $F_{n}$ is the $n$th Fibonacci number.
\end{thebibliography}
\end{document}